\documentclass[twocolumn,aps,groupedaddress]{revtex4}

\usepackage{graphicx}
\usepackage{dcolumn}
\usepackage{bm}
\usepackage{amsmath,amssymb,amsfonts}
\usepackage{float}
\usepackage{color}
\usepackage{dcolumn}
\usepackage{wasysym}
\newcolumntype{N}{@{}m{0pt}@{}}

\begin{document}

\title{Electronic structures, charge transfer and charge orders in twisted transition metal dichalcogenide bilayers}

\author{Yang Zhang}
\author{Tongtong Liu}
\author{Liang Fu}
\affiliation{Department of Physics, Massachusetts Institute of Technology, Cambridge, Massachusetts 02139, USA}

\begin{abstract}
Moir\'e superlattices of transition metal dichalcogenide (TMD) bilayers have been shown to host correlated electronic states, which arise from the interplay of long wavelength moir\'e potential and long-range Coulomb interaction.
Here we theoretically investigate structural relaxation and single-particle electronic structure of twisted TMD homobilayer. From the large-scale density functional theory calculation and continuum model with layer degrees of freedom, we find that the out-of-plane gating field creates a tunable charge transfer gap at the Dirac point between the first and second moir\'e valence bands. We further study the charge orders at the fractional band fillings. In the flat band limit, we find from Monte Carlo simulations a series of charge-ordered insulating states at various fillings $n=1/4, 1/3, 1/2, 2/3, 1$. We predict that gating field induces a phase transition between different electron crystals at fixed filling $n=1/2$ or $2/3$. At half-filling $n=1$, the ground state is a Mott insulator with electronically driven ferroelectricity. Our work demonstrates that transition metal dichalcogenide homobilayer provides a powerful platform for the investigation of tunable charge transfer insulator and charge orders.


\end{abstract}

\maketitle

Moir\'e superlattices are a fruitful platform for realizing and controlling correlated electron states, as evidenced by the remarkable success in twisted bilayer graphene (TBG) \cite{cao2018correlated,cao2018unconventional,lu2019superconductors,kerelsky2019maximized,jiang2019charge,xie2019spectroscopic,choi2019electronic,yankowitz2019tuning,codecido2019correlated,sharpe2019emergent,tomarken2019electronic,zondiner2019cascade} and trilayer graphene-hBN heterostructure\cite{chen2019evidence,chen2019sig,serlin2019intrinsic,liu2019spin}. Recently a new family of moir\'e materials based on transition metal dichalcogenides (TMD) \cite{zhang2017interlayer,yu2017moire,seyler2019signatures,tran2019evidence,yuan2020twist,li2020dipolar,brotons2020spin,shimazaki2020strongly,bai2020excitons,wang2020correlated,mcgilly2020visualization,zhang2020flat,weston2020atomic} have attracted great interest. They host an abundance of correlated insulating states at a series of fractional fillings \cite{regan2019optical,tang2019wse2,xu2020abundance,huang2020correlated,jin2020stripe}.


In TMD bilayers, moir\'e bands are formed from parabolic bands of individual layers. In twisted TMD homobilayers, the moir\'e bandwidth can be made arbitrarily small by reducing the twist angle, which gives rise to strong correlation without fine tuning. Electrons or holes in these moir\'e bands are tightly localized in high-symmetry stacking regions, which can be well described by a simple effective tight-binding model. This description offers a convenient starting point for investigating interaction-induced states at finite density.
Despite the conceptual simplicity, a quantitative modeling of moir\'e bands in TMD is highly nontrivial. For example, the moir\'e bandwidth of TMD heterobilayer WSe$_2$/WS$_2$ is only on the order of 10 meV, and depends highly on the lattice relaxation \cite{regan2019optical,tang2019wse2,zhang2020moire,li2020imaging}.

In this work, using the large-scale density functional theory, continuum model approach and Monte Carlo simulation, we study the effect of structural relaxation and electric field on the moir\' e band structure in twisted TMD homobilayers and predict novel charge orders at fractional fillings in the strong-coupling regime. We focus on the moir\'e valence bands originating from the $\Gamma$ pocket \cite{naik2018ultraflatbands,xian2020realization,angeli2020gamma,zhan2020multi,venkateswarlu2020electronic}.
Due to interlayer tunneling and lattice relaxation, these moir\'e bands are derived from localized orbitals in MX and XM stacking regions that form a honeycomb lattice. We find a pair of massless Dirac fermions at $K, K'$ points of the mini Brillouin zone (BZ), which is protected by the $D_3$ point group symmetry of the moir\'e superlattice. Applying an out-of-plane electric field breaks the sublattice symmetry of the honeycomb lattice and opens a tunable gap $\Delta$ at the Dirac point. We introduce a new continuum model for twisted TMD homobilayers, which captures the layer degrees of freedom and the electrically tunable gap.

We further use an extended Hubbard model on the honeycomb lattice and perform Monte Carlo simulations to study the insulating electron crystals in the flat band limit. 
We find a distinctive set of charge orders at hole fillings $n=1/4, 1/3, 1/2, 2/3, 1$ on the honeycomb lattice.
Interestingly, the charge orders at $n=1/2$ and $2/3$ both break the rotational symmetry and differ from the proposed states in the WSe$_2$/WS$_2$ heterobilayer. And the $n=1$ insulating state has a spontaneous out-of-plane ferroelectric polarization, which can be switched by the electric field.
These symmetry breaking charge orders can be directly probed by the optical anisotropy experiments \cite{jin2020stripe,shimazaki2020optical}. Moreover, we predict that phase transitions between distinct charge-ordered states at the same filling can be induced by the electric field, which tunes the charge-transfer gap $\Delta$. Our work shows that twisted homobilayer MoS$_2$ provides an ideal platform for investigating electrically tunable charge transfer gap and charge orders.

We study TMD homobilayers with a small twist angle starting from AA stacking, where every metal (M) or chalcogen (X) atom  on  the top layer is aligned with the same type of atom on the bottom layer \footnote{AB stacking can be viewed as a $180^{\circ}$ rotation of top layer}. 
Within a local region of a twisted bilayer, the atom configuration is identical to that of an untwisted bilayer, where one layer is laterally shifted relative to the other layer by a corresponding displacement vector ${\bm d}_0$. For this reason, the moir\'e band structures of twisted TMD bilayers can be constructed from a family of untwisted bilayers at various ${\bm d}_0$, all having $1\times 1$ unit cell. Our analysis thus starts from untwisted bilayers \cite{mcdonnell2014hole}.

\begin{figure}[t]
\includegraphics[width=1.0\linewidth]{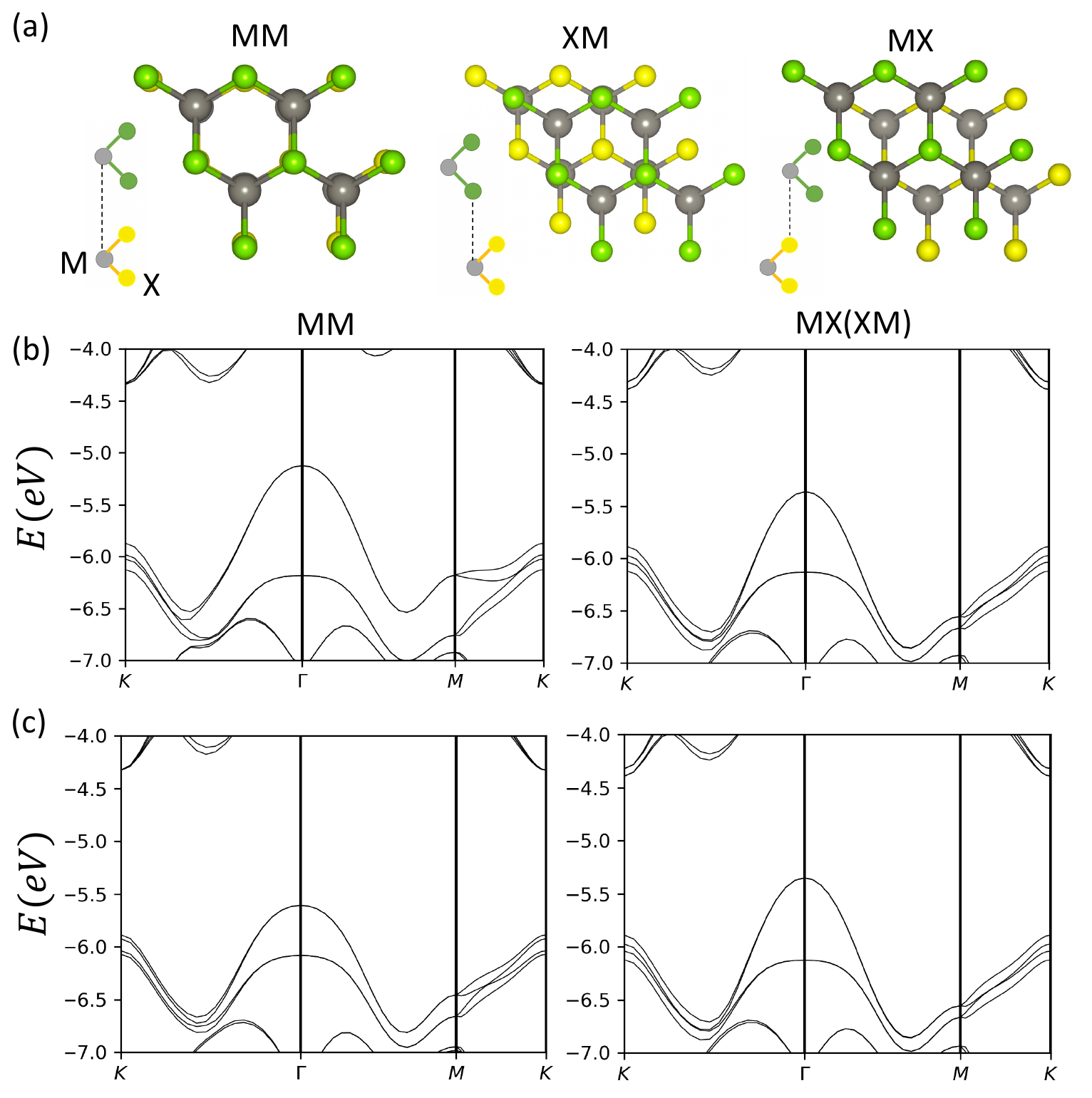}
\caption{
(a) Lattice structure of MM, MX, XM spots for AA stacking heterobilayer, M stands for metal atom and X stands for chalcogen atom (Green for the top layer, yellow for the bottom layer).
DFT band structures of MM and MX(XM) stacking homobilayer in
(b) MoS$_2$/MoS$_2$ with identical layer spacing;
(c) MoS$_2$/MoS$_2$ with relaxed layer spacing.
}
\label{fig1}
\end{figure}

In particular, ${\bm d}_0=0, -\left({\bm a}_{1}+{\bm a}_{2}\right) /3, \left({\bm a}_{1}+{\bm a}_{2}\right) /3$, where ${\bm a}_{1,2}$ is the primitive lattice vector  for untwisted bilayers, correspond to three high-symmetry
stacking configurations of untwisted TMD bilayers, which we refer to as MM, XM, MX. In MM (MX) stacking, the M atom on the top layer is locally aligned with the M (X) atom on the bottom layer, see Fig. \ref{fig1}a. Likewise for XM. The bilayer structure in these stacking configurations is invariant under three-fold rotation around the $z$ axis.

In homobilayer TMD, the spin degenerate $\Gamma$ pockets in the valence band arise from electron tunneling between the two layers. The $k\cdot p$ Hamiltonian takes the form:
\begin{equation}
\mathcal{H}\left( \boldsymbol{d}_{0}\right)=\left(\begin{array}{cc}
-\frac{\hbar^{2} k^{2}}{2 m^{*}}+\epsilon_{b}\left(\boldsymbol{d}_{0}\right)  & \Delta_{T}\left(\boldsymbol{d}_{0}\right) \\
\Delta_{T}^{\dagger}\left(\boldsymbol{d}_{0}\right) & -\frac{\hbar^{2} k^{2}}{2 m^{*}}+\epsilon_{t}\left(\boldsymbol{d}_{0}\right)
\end{array}\right).
\end{equation}
Here $m^*=1.07m_e$ is the effective mass for the valence band. $\Delta_T\left(\boldsymbol{d}_{0}\right)$ is the interlayer tunnelling amplitude which depends on the in-plane displacement between the two layers. In contrast to the complex tunneling amplitude for the $K$ pockets \cite{wu2019topological}, here the time-reversal symmetry at $\Gamma$ pocket enforces $\Delta_T\left(\boldsymbol{d}_{0}\right)$ to be  real. The potential term $\epsilon_{b,t}\left(\boldsymbol{d}_{0}\right)$ denotes the energy of the valence band maximum in the absence of tunneling, which arises from the unequal layer weight of the wavefunction at MX and XM stacking configuration.

We expand $\Delta_T\left(\boldsymbol{d}_{0}\right)$ in Fourier components up to the second harmonic term:
\begin{equation}
\Delta_{T}\left(\boldsymbol{d}_{0}\right)=
w_0+
2w_1 \sum^3_{j=1} \cos( \boldsymbol{G}_{j} \cdot \boldsymbol{d}_{0})+
2w_2 \sum^3_{j=1} \cos( \boldsymbol{2G}_{j} \cdot \boldsymbol{d}_{0}), \label{Dt}
\end{equation}
where $G_i(i=1,2,3)$ are the three reciprocal lattice vectorc in monolayer TMD. Due to three-fold rotational symmetry, $\Delta_T$ is a local extremum for MM, MX and MX stackings, with $\Delta_T=w_0+6w_1+6w_2$ for $d_0$=0 (MM) and $w_0-3w_1-3w_2$
for $d_0= \pm\left({\bm a}_{1}+{\bm a}_{2}\right) /3$ (MX or XM).
The zero-momentum-transfer tunneling term $w_0$ is responsible for the large bonding and antibonding energy splitting for all $d_0$, while $w_1, w_2$ capture the variation of the tunneling amplitude at different lateral displacements.

The interlayer tunneling strength depends significantly on the layer spacing $d$. 
From the DFT calculation, we find the equilibrium layer spacing of untwisted TMD bilayers in MM, MX and XM stackings: $d_{MM}=6.63$ Angstroms and $d_{MX}=d_{XM}=5.97$ Angstroms. The 10\% variation of layer spacing is comparable with that in bilayer graphene \cite{uchida2014atomic} and strongly impacts the energy splitting of $\Gamma$ pockets.  

By calculating the work function, we plot in Fig. \ref{fig1} the band structure of MM and MX-stacked bilayers, with reference energy $E=0$ chosen to be the absolute vacuum level. Using the relaxed layer spacings, we find the energy splitting in MX (or XM) stacking to be stronger than in $MM$, as a result of its smaller layer distance. From the different energy splitting at Fig. \ref{fig1}c, we obtain the tunnelling parameters as $w_0=338$ meV, $w1+w2=-18$ meV. If the same layer spacing were used for both MX and MM bilayers, the opposite (and incorrect) conclusion about the energy splitting would be found, see Fig.\ref{fig1}b. Thus lattice relaxation is crucial for obtaining the correct moir\'e band structure.

\begin{figure}[t]
\includegraphics[width=1.05\linewidth]{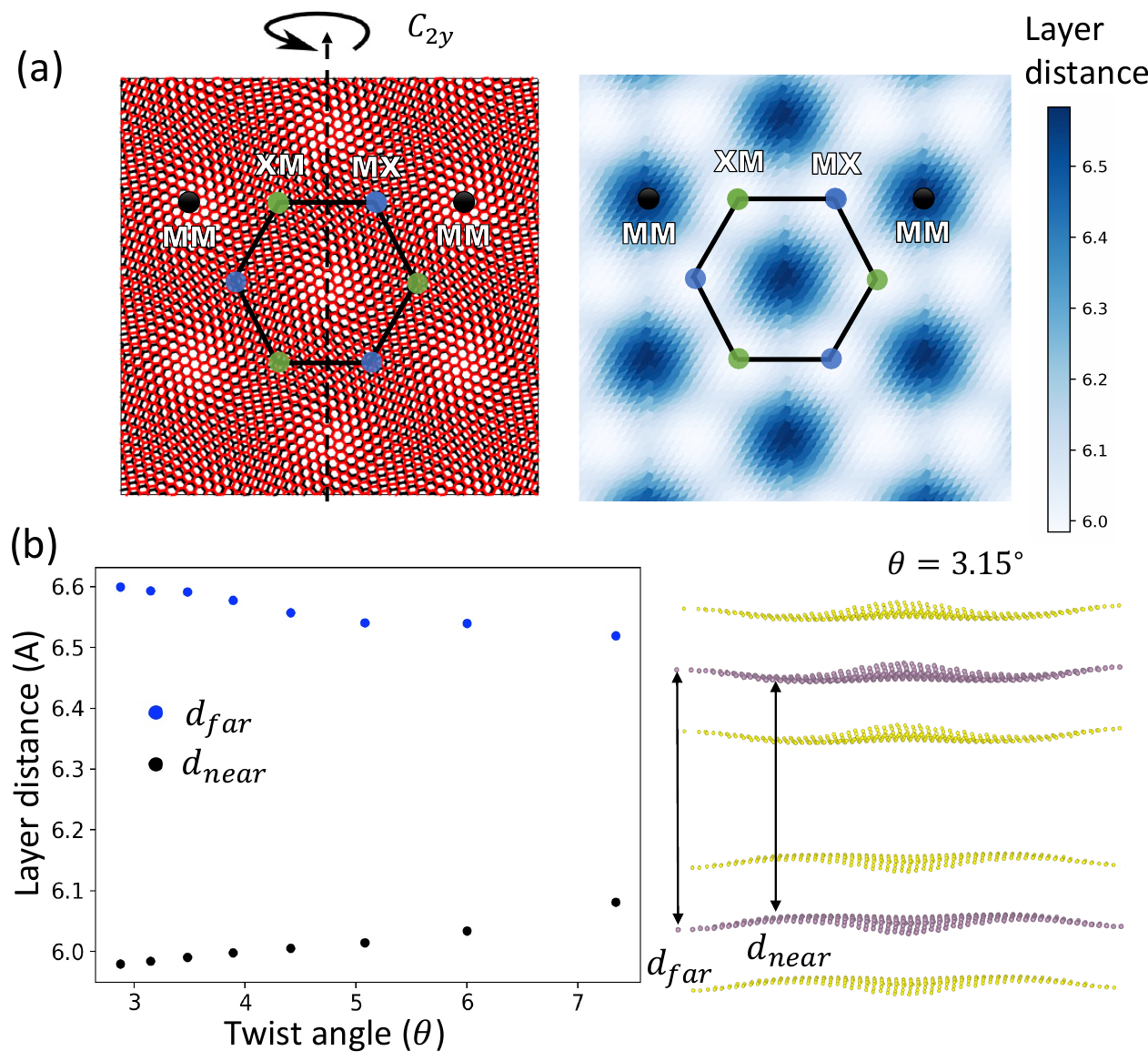}
\caption{
(a) Real-space moir\'e pattern of heterobilayer TMD heterobilayer, where MM, MX, XM spots within one supercell are labeled, and the diagram for spacial dependent layer distance (in the unit of Angstrom) in the moir\'e superlattice;
(b) Twist-angle dependent layer spacing for $d_{far}$ and $d_{near}$, and out of plane corrugation.
}
\label{fig2}
\end{figure}


The structure of twisted TMD homobilayers can be described by a lateral shift ${\bm d}_0$ that varies slowly in space: ${\bm d}_0 = \theta \hat{z} \times \boldsymbol{r}$. 
Therefore we construct the following continuum Hamiltonian for the moir\'e bands from $\Gamma$ pocket two-band $kp$ model:
\begin{equation}
\mathcal{H}=\left(\begin{array}{cc}
-\frac{\hbar^{2}k^{2}}{2 m^{*}}+\epsilon_{b}(r) & \Delta_{T}(r) \\
\Delta_{T}^{\dagger}(r) & -\frac{\hbar^{2}k^{2}}{2 m^{*}}+\epsilon_{t}(r)
\end{array}\right)
\end{equation}
The position-dependent tunneling term is obtained by replacing ${\bm d}_0$ with $\theta \hat{z} \times \boldsymbol{r}$ in Eq.(\ref{Dt}):
\begin{equation}
\Delta_{T}\left(\boldsymbol{r}\right)=
w_0+
2w_1 \sum^3_{j=1} \cos( \boldsymbol{G^m_j} \cdot \boldsymbol{r})+
2w_2 \sum^3_{j=1} \cos( \boldsymbol{2G^m_j} \cdot \boldsymbol{r})
\end{equation}
Where $\boldsymbol{G^m_i}= \boldsymbol{G_i}\theta\times \hat{z}  (i=1,2,3)$ are the three reciprocal lattice vectors in moir\'e superlattice. Likewise, the intralyer potential $\epsilon_{t,b}$ ($t, b$ stand for top and bottom layer, respectively) can be expressed as the first order Fourier expansion over moir\'e reciprocal lattice vector:
\begin{equation}
\epsilon_{t,b}\left(\boldsymbol{r}\right)=2V_0 \sum_{j=1,2,3} \cos \left(\boldsymbol{G^m_j} \cdot \boldsymbol{r} \pm \phi\right)
\end{equation}
The sign of phase factor $\phi$ changes under layer exchange, enforced by $C_{2y}$ symmetry as shown in Fig. \ref{fig2}a. The potential term is crucial for the later modelling with out-of -plane gating field.

We now compare the band structure from continuum model with the large scale density functional theory. The moir\'e superlattice is fully relaxed with van der Waals correction incorporated by the vdW-DF (optB86) functionals \cite{klimevs2011van} as implemented in the Vienna Ab \textit{initio} Simulation Package\cite{kresse1996efficiency}.
We plot the twist-angle dependent layer distance, $d_{far}$ at MM region, and $d_{near}$ in MX (XM) region, in Fig. \ref{fig2}b.
At small twist angle $\theta \sim 0$, the two layers are corrugated,
and the layer distance of MM, MX or XM stacking region approaches to that of the untwisted structure. The interlayer tunneling amplitude is maximum at MX and XM regions, which are related by $C_{2y}$ symmetry. As a result, low-energy moir\'e bands are formed from layer-hybridized orbitals in MX and XM regions, which form a honeycomb lattice with identical on-site potential.

\begin{figure}[t]
\includegraphics[width=1.0\linewidth]{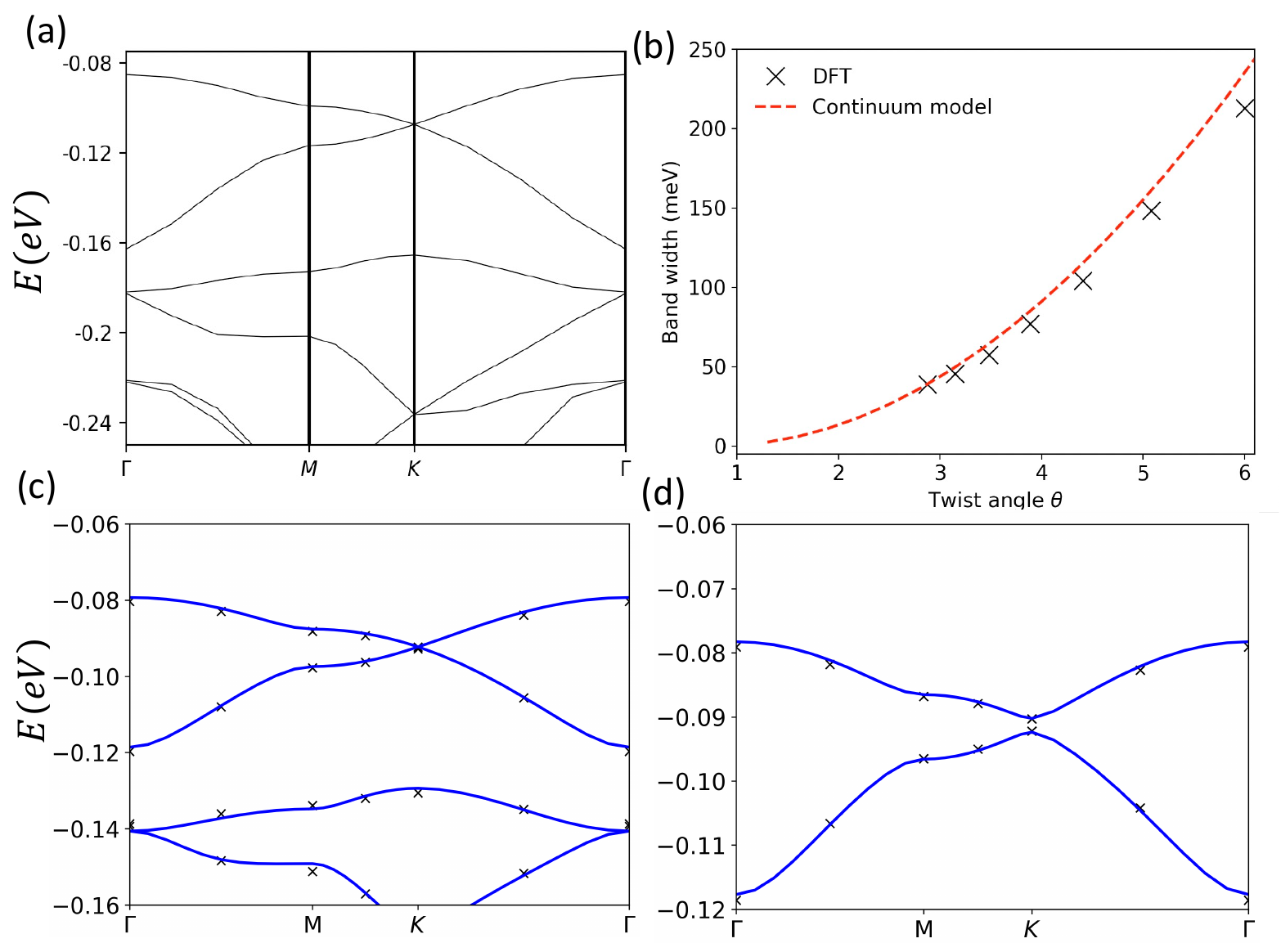}
\caption{
(a)DFT Band structure for  $\theta=3.89^\circ$;
(b)Twist angle dependent bandwidth for the first two moir\'e bands (top two valence bands in (a) ) of the honeycomb lattice.
DFT (black cross) and continuum model(blue line) band structures for
(c) $\theta=2.876^\circ$; (d) $\theta=2.876^\circ$ with 0.5 $V/nm$ out of plane gating field.
}
\label{fig3}
\end{figure}

We perform the large scale DFT simulation to calculate the band structures for various twist angles, shown in Fig. \ref{fig3}. We find that above a small moir\'e period $L_m \sim 4.7$ nm with twist angle $\theta=3.89^\circ$, the two topmost moir\'e $s$ bands are well separated from the remaining bands. Similar band structures are also found in large-scale DFT calculation with fully relaxed lattice structure for homobilayer MoS$_2$\cite{naik2018ultraflatbands,xian2020realization} and WS$_2$\cite{angeli2020gamma}.
Fitting the DFT moir\'e band structure to continuum model, we obtain the parameters as $w_0=338$ meV, $w_1=-16$ meV and $w_2=-2$ meV, $V_0=6$ meV, $\phi=121^\circ$ at twist angle $\theta=2.876^\circ$. These values are consistent with the estimation from untwisted structures.


As shown in Fig. \ref{fig3}(a,c), the moir\'e bands exhibit Dirac points at $K$ and $K'$ points of the moir\'e Brillouin zone. These Dirac points are protected by the $D_3$ point group of twisted TMD homobilayer: the doublet at $K$ or $K'$ form a two-dimensional $E$ representation. The bandwidth of Dirac bands changes monotonously from 250 meV to 10 meV when twist angle $\theta$ ranges from $6^\circ$ to $2^\circ$ as shown in Fig. \ref{fig3}b. This provides an ideal platform to study the tunable correlation physics of Dirac electrons at the filling of $n=2$ per moir\'e unit cell.

In the case of twisted bilayer graphene \cite{bistritzer2011moire}, the low energy Dirac fermion is protected by the $C_{2z}$ symmetry, which can not be broken by the out-of-plane field. However, in MX and XM regions of the twisted homobilayer MoS$_2$, the wavefunctions have unequal layer weight as indicated from the untwisted calculation. Thus the out-of-plane gating field breaks the $C_{2y}$ symmetry and gaps out the Dirac fermion. A simplified continuum model targeting at antibonding orbitals well captures the topmost moir\'e bands, but can not describe the band structure and charge distribution involving layer degrees of freedom.

We further calculate the band structure of the fully relaxed moir\'e superlattice of homobilayer MoS$_2$ with the applied gating field.
As shown in Fig. \ref{fig3}d, an out-of-plane gating field 0.5 $V/nm$ creates a 2.4 meV gap at K point, while the bandwidth of the first energy-separable moir\'e band is 12 meV. At $K$ point of the band edge, the wavefunction of the first band is localized at MX region, while the second band at XM region. For small twist angle $\theta=2^\circ$ with wavelength $L_m=9.1$ nm, the gating field $E_d=1 V/nm$ induces a charge transfer gap $\Delta$ up to 5 meV, even larger than the bandwidth of the topmost moir\'e band (see supplementary material). A larger field-induced $\Delta$ can be achieved in twisted TMD homobilayers with reduced interlayer tunneling (which competes with the layer potential asymmetry). This can be realized by inserting an hBN layer in between the top and bottom TMD layers \cite{shimazaki2020strongly}.



In the TMD superlattice, the local minimums of the periodic moir\'e potential can be viewed as the effective moir\'e atoms to host charge. Under the harmonic approximation, the size of the Wannier orbital for the topmost moir\'e band is given by $\xi=\sqrt{\frac{\hbar}{m^* \omega}}=2\left( \pi\right)^{-\frac{1}{2}} \sqrt{L_{m}}(\frac{\hbar^2}{m^*V_m})^{\frac{1}{4}}$ \cite{zhang2020moire}($V_m$ is the moir\'e potential integrated to antibonding orbitals). In homobilayer system without lattice mismatch, the kinetic energy over nearest neighbor interaction ($t/V_1$) can be tuned arbitrarily small, so that the classical model is well justified at sufficiently small twist angle. The effective extended Hubbard model without kinetic energy is given by:
\begin{equation}
\begin{array}{rlrl}
H_{0}  =\sum_{j \in B} \Delta n_{j}+\sum_{i} U n_{i \uparrow} n_{i \downarrow}+ & \frac{1}{2} \sum_{i \neq j} V_{i j} n_{i} n_{j}
\end{array}
\end{equation}
Here $\Delta$ is the charge transfer gap between two sublattice sites A and B, and $V_{ij}$ is the extended interaction between $i$ and $j$ sites.

\begin{figure}[t]
\includegraphics[width=1.0\linewidth]{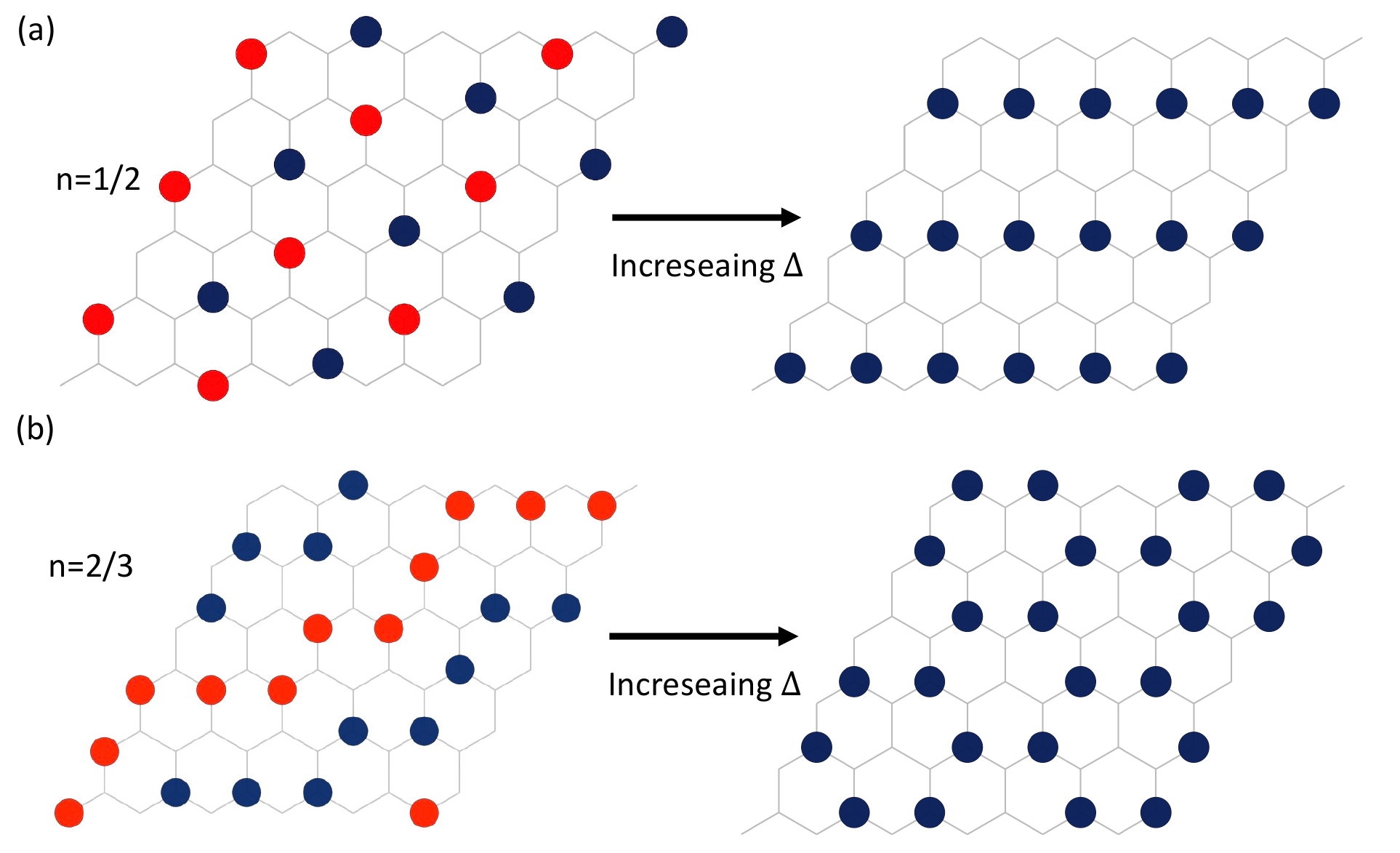}
\caption{
Ground state charge order at filling
(a) n=1/2 with increasing charge transfer gap $\Delta$, (b) n=2/3 with increasing charge transfer gap $\Delta$.
}
\label{fig4}
\end{figure}

In twisted homobilayer MoS$_2$, the gating field introduces a charge transfer gap $\Delta$. We first discuss the situation with large $\Delta$. At filling $n<1$, the effective tight-binding model reduces to a triangular lattice model, as in the case of WSe$_2$/WS$_2$, and exhibits similar charge orders. Various insulating states have been observed at fractional fillings $n=1/4, 1/3, 2/5, 1/2, 3/5, 2/3$\cite{xu2020abundance,regan2019optical,tang2019wse2}. Due to the strong on-site Coulomb repulsion $U>>\Delta$, the system at $n=1$ should be regarded as a charge transfer insulator \cite{zhang2020moire}. When doped to a higher filling $n>1$, additional charges transfer to the other sublattice/layer.

Here we further study the charge orders of honeycomb lattice with small $\Delta$ including $\Delta=0$ in flat band limit. We perform classical Monte Carlo simulation up to $120\times 120$ sites with periodic boundary condition for the extended Hubbard model with different gate distances from $d=L_m/2$ to $d=10L_m$. The distance dependent interaction strength is plotted in Fig. S2 up to $V_{100}$, and the interaction cutoff is chosen as $0.1\%V_1$. We identify a series of charge orders at $n=1/4, 1/3, 1/2, 2/3, 1$. For $n<1/2$, moir\'e electrons are all filled to one sublattice, exhibiting similar charge orders (or generalized Wigner crystals) as observed in WSe$_2$/WS$_2$ heterobilayer \cite{padhi2018doped,regan2019optical,padhi2021generalized}.

Interestingly, for small $\Delta$, we find that charge transfer involving two sublattices already takes place for filling $n\ge 1/2$, leading to new charge-ordered states beyond those found in  WSe$_2$/WS$_2$. At filling factor $n=1/2$, we find an emerging rectangular lattice with $\sqrt{3}\times 2$ periodicity. This state breaks the three-fold rotational symmetry and can be viewed as the combination of the stripe states on both sublattices, each at $1/4$ filling. This rectangular electron crystal is energetically favorable compared to the enlarged $2\times 2$ honeycomb crystal at all gate screening distances. In contrast, at large $\Delta$, the ground state becomes a simple stripe state on the triangular sublattice with lower on-site potential, as shown in Fig. \ref{fig4}. We find the critical charge transfer gap is $\Delta_c=2(V_2-V_3-V_4+2V_6-V_9+V_{12}+...)$. For $d=L_m=9.1$ nm, $\Delta_c=0.12 \frac{e^2}{\epsilon L_m} \sim 3.8$ meV can be reached by realistic gating field. We note the critical $\Delta_c$ can be further lowered by increasing moir\'e wavelength.


At filling factor $n=2/3$, the charges form a zigzag stripe order with $6\times 6$ periodicity breaking the $C_3$ rotational symmetry. This zigzag type charge configuration is energetically favored compared to a linear stripe at screening distances from $d=1/2L_m$ to $d=10L_m$. As $\Delta$ increases, the zigzag charge stripe transitions to the $\sqrt{3}\times\sqrt{3}$ crystal that occupies one sublattice sites only, as shown in Fig. \ref{fig4}. We find the critical charge transfer gap is $\Delta_c=V_2-V_3-\frac{10}{3}V_4+\frac{14}{3}V_5....=0.04 \frac{e^2}{\epsilon L_m} \sim 1.3$ meV at $d=L_m=9.1$ nm.

The transition between distinct electron crystals at the same filling is first-order. This should lead to a kink in the sublattice/layer charge imbalance as a function of the gating field.
This prediction, which is a main result of our work, can be tested in MoSe$_2$/hBN/MoSe$_2$ heterostructure \cite{shimazaki2020strongly}, where the gating field induced charge transfer between the top and bottom layers has already been observed at relatively high temperature. 

For $n=1$, we find that even at $\Delta=0$, the ground state is a fully sublattice polarized Mott insulating state, which spontaneously breaks the honeycomb lattice symmetry. As the two sublattice sites MX and XM have different layer weight, the Mott insulating state at n=1 develops a finite out-of-plane ferroelectric polarization, which can be switched by the electric field. The ferroelectricity driven by the Mott physics in TMD moir\'e systems goes beyond the conventional ferroelectricity and enable the fast switching due to electronic origin \cite{zheng2020unconventional}.
For filling $n>1$, charge-2e trimer can be the lowest energy excitation when tuning the charge transfer gap $\Delta$, providing a platform to design unconventional superconductivity \cite{slagle2020charge}.

In homobilayer WSe$_2$, the valence band maximum is located at $K$ with weak interlayer tunneling amplitude and intralayer potential both on the order of 10 meV. The complex tunneling term between two layers brings further complications for the theoretical and experimental investigation of the insulating states \cite{wu2019topological,wang2020correlated,zhang2020flat}. 

In conclusion, we present a combined study of lattice relaxation, single-particle electronic structure, and ground state charge orders on the twisted homobilayer MoS$_2$. Unlike the previous moir\'e charge transfer insulator in WSe2$_2$/WS$_2$ heterobilayer, here out-of-plane gating field breaks $C_{2y}$ symmetry and induces a controllable charge transfer gap. With Monte Carlo simulation, we predict additional stripe-type charge orders at fillings $n=1/2, 2/3$ in the emergent honeycomb lattice with $\Delta=0$. When increasing $\Delta$, these electron crystals transit to fully sublattice polarized states. We further predict the ferroelectricity at the $n=1$ Mott insulating state, which enables the ultrafast switching of electronic polarization. Our work demonstrates that the interplay between two moir\'e regions leads to the charge transfer insulator \cite{zaanen1985band,zhang2020moire} and serves as a platform for creating novel correlated states, such as unconventional density wave \cite{pan2020band,pan2020quantum}, charge stripes \cite{jin2020stripe}, spin superfluid\cite{bi2019excitonic} and superconductivity\cite{slagle2020charge}.

\section*{Acknowledgment}
We thank Atac Imamoglu, Yuya Shimazaki, Cory R. Dean and Qianhui Shi for numerous discussions on experiments, and Zhen Bi, Lede Xian, and Angel Rubio for valuable theoretical discussions. This work is supported by DOE Office of Basic Energy Sciences, Division of Materials Sciences and Engineering under Award DE-SC0018945. L.F. is partly supported by a Simons Investigator award from the Simons Foundation.



\newpage

\bibliography{ref}

\end{document}